\documentclass[11pt]{article}  % DEFINIÇÃO QUE RESOLVEU  O  PROBLEMA COM A  FIGURA

\usepackage{graphicx,color,epsf} % PACOTE PARA INSERIR FIGURAS

\usepackage{caption}
\begin{document}

%\begin{flushright}
%QMW/95-5
%\\hep-th/9502016
%\end{flushright}
%\vspace{0.5cm}
\begin{center}
{\LARGE \bf Stefan-Boltzmann law for massive photons}
\\
\vspace{1cm}
{\large  E. S. Moreira Jr.$^{a}$ and T. G. Ribeiro}$^{b,}$\footnote{E-mail:
moreira@unifei.edu.br, tiaggribeiro@gmail.com.}
\\
\vspace{0.3cm}
{$^{a}$\em Instituto de Matem\'{a}tica e Computa\c{c}\~{a}o,}
{\em Universidade Federal de Itajub\'{a},}   \\
{\em Itajub\'a, Minas Gerais 37500-903, Brazil}  \\
\vspace{0.1cm}
{$^{b}$\em Instituto de Ci\^{e}ncias Exatas,}
{\em Universidade Federal de Juiz de Fora,}   \\
{\em Juiz de Fora, Minas Gerais 36036-900, Brazil}

%\vspace{0.3cm}
%{\large August, 2015}
\end{center}
%\vspace{0.5cm}

\begin{abstract}
This paper generalizes the Stefan-Boltzmann law to include massive photons.
A crucial ingredient to obtain the correct formula for the radiance is to realize
that a massive photon does not travel at the speed of (massless) light.
It follows that, contrary to what could be expected, the radiance
is not proportional to the energy density times the speed of light.

%Thirty years ago a paper appeared in the literature
%generalizing the Stefan-Boltzmann law to include massive photons.
%Torres-Hern\'{a}ndez suggested
%in [Phys. Rev. A {\bf 32}, 623 (1985)] to check 
%the mass of the photon
%has nonvanishing mass
%by determining modifications of the
%Stefan-Boltzmann law due to a nonvanishing mass. 
%The paper suffers from a flaw though: 
%it assumes that a massive photon
%travels at the speed of (massless) light. The present work fixes the
%mistake and presents the correct formula for the radiance.  
\end{abstract}
%\vspace{1cm}

\section{Introduction}

The understanding of the the Stefan-Boltzmann law is one of the most celebrated landmarks of modern physics. In order to derive it an important aspect is the speed at which
energy leaves the blackbody cavity. If one considers
the electromagnetic radiation composed of massless photons, the speed they leave the cavity is taken to be the speed of light c,
leading to a formula for the radiance as being proportional to the energy density times $c$, i.e., the familiar Stefan-Boltzmann law \cite{lan80}.

The present work addresses this issue  for a gas of massive photons.
Formulas for the mean speed $\bar{v}$ and the radiance ${\cal R}$
are determined for arbitrary mass of the photon and at all temperatures.
%The argument in Ref. \cite{tor85} will be amended, 
The corresponding high temperature (or small mass) and  low temperature 
(or large mass) regimes are discussed.
It should be noted that the Stefan-Boltzmann law for massive photons was considered previously in the literature \cite{tor85}. However, the calculations were spoiled by the incorrect assumption that the massive photon's speed is $c$.

%In the limit of small mass,  ${\cal R}$ will be compared with Stefan- %Boltzmann's law. 
%It should be mentioned that the error in
%Ref. \cite{tor85} is propagating in the literature (including a review, Ref. \cite{tu05})  %stressing the pertinence of this work. 

\section{Speed of a massive photon}

As is well known there are two equivalent ways to arrive at the
thermodynamics of radiation at temperature $T$ in a cavity of volume $V$.
One is to consider an ensemble of quantum harmonic oscillators, 
and another that treats hot radiation as a Bose gas of relativistic particles
with vanishing chemical potential \cite{lan80}.
According to the latter approach the distribution function is given
by 
\begin{equation}
f({\bf p})=\frac{2}{e^{\beta \epsilon}-1},
\label{distribution}
\end{equation}
where $\beta:=1/k T$ and $\epsilon:=pc$ is the energy of the photon
of momentum ${\bf p}$ (note that $p:=||{\bf p}||$). Now, assuming that the photon
has mass $m$ the expression for $\epsilon$ is replaced by
\begin{equation}
\epsilon:=\sqrt{p^2 c^2+m^2 c^4}.
\label{energy}
\end{equation}
In fact, the factor 2 in  Eq. (\ref{distribution}) 
should be replaced by 3 taking into account an additional degree of freedom 
associated with a nonvanishing $m$; however it will be assumed that
such a degree of freedom can be ignored in the thermal equilibrium
of radiation and matter (see e.g. Ref. \cite{tor85}).

The number of photons $N$ in the cavity is obtained  by integrating
Eq. (\ref{distribution}) over the phase space of one photon,
\begin{eqnarray}
&&\frac{N}{V}=\int \frac{d^3 p}{h^3}f({\bf p})=
\frac{8\pi}{h^3}\int_{0}^{\infty}\- dp\ \frac{p^2}{e^{\beta\epsilon}-1}, 
\label{number}
\end{eqnarray}
where spherical coordinates have been used in the last equality.
With the help of the geometric series and noting Eq. (\ref{energy}),
Eq. (\ref{number}) can be recast as
\begin{eqnarray}
&&\frac{N}{V}=
\frac{8\pi}{(hc)^3}\sum_{n=1}^{\infty}
\int_{mc^2}^{\infty}d\epsilon\ \epsilon\sqrt{\epsilon^2 -m^2c^4}\ 
e^{-n\beta\epsilon}, 
\label{geometricseries}
%\label{number2}
\end{eqnarray}
where one identifies  an integral representation of the modified Bessel function of the second kind $K_{2}(z)$, resulting in
\begin{eqnarray}
&&\frac{N}{V}=
\frac{m^2 c kT}{\pi^2\hbar^3}
\sum_{n=1}^{\infty}n^{-1}K_{2}\left(n x \right),
\hspace{1.5cm}
x:=\frac{mc^2}{kT} .
%\nonumber
\label{number2}
\end{eqnarray}
As a simple check,
considering that  for very small arguments 
$K_{2}(z)$ behaves approximately as $2/z^2$, by setting $m\rightarrow 0$
in Eq. (\ref{number2}) it follows 
$N/V=(2\zeta(3)/\pi^2)(kT/\hbar c)^3$ which is a familiar expression of blackbody radiation. 

\begin{figure}[h]
\center
\includegraphics[scale=0.40]{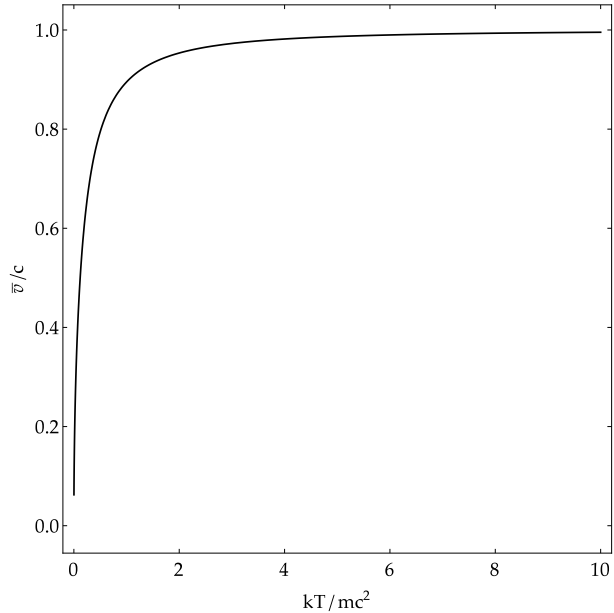}
%{graficofinal.eps}
%\includegraphics[scale=0.40]{grafico1.eps}
\caption{\footnotesize{Thermal behavior of
 the mean speed $\bar{v}$ of a massive photon according to Eq. (\ref{photonspeed}). Note that as $T\rightarrow\infty$, 
 $\bar{v}\rightarrow c$, and that at low temperatures $\bar{v}$  behaves
approximately as in Eq. (\ref{idealgas}).}}
\label{pressaobessel}
\end{figure}

The speed of the photon ${\bf v}$ and its momentum are related by
${\bf p}=(\epsilon/c^2){\bf v}$, leading to (below $v:=||{\bf v}||$)
\begin{eqnarray}
&&
p=\frac{\epsilon}{c^2}v,
\hspace{1.5cm}
v=c\ \sqrt{1-(mc^2/\epsilon)^2},
\label{speed}
\end{eqnarray}
clearly showing that $v=c$ only if the photon is massless.
The mean value of $v$ can be obtained by averaging over
the phase space with Eq. (\ref{distribution}), i.e.,
\begin{eqnarray}
&&\bar{v}= \frac{V}{N}\int \frac{d^3 p}{h^3}vf({\bf p})=
\frac{8\pi c^2 V}{N h^3}\int_{0}^{\infty}\- dp\ \frac{p^3}{\epsilon(e^{\beta\epsilon}-1)}, 
\label{speed2}
\end{eqnarray}
where the  first expression in Eq. (\ref{speed}) has been used.
Now, following along the same steps that led from Eq. (\ref{number}) to Eq. (\ref{number2}) and using the formula for $N$ in Eq. (\ref{number2}), 
one finds,
\begin{equation}
\bar{v}=2c 
\left[Li_{3}(e^{-x})+xLi_{2}(e^{-x})\right]
\left[x^2\sum_{n=1}^{\infty} n^{-1}K_{2}(nx)\right]^{-1},
%\frac{Li_{3}(e^{-x})+xLi_{2}(e^{-x})}
%{x^2\sum_{n=1}^{\infty} n^{-1}K_{2}(nx)}
\label{photonspeed}
\end{equation}
where 
\begin{equation}
Li_{s}(z):=\sum_{n=1}^{\infty} n^{-s}z^n
\label{poly}
\end{equation}
%$Li_{s}(z):=\sum_{n=1}^{\infty} n^{-s}z^n$ 
is the polylogarithm
function [note that $Li_{s}(1)=\zeta(s)$].
By taking $m\rightarrow 0$ [i.e., $x\rightarrow 0$, cf. Eq. (\ref{number2})] in Eq. (\ref{photonspeed}) it results $\bar{v}=c$ for all values of $T$. However, if $m$ is strictly nonvanishing, when $x\gg 1$ (i.e., at low temperatures),
Eq. (\ref{photonspeed}) yields the following asymptotic behavior
[recall that $K_{\nu}(z)$ behaves approximately as $\sqrt{\pi/2z}\ e^{-z}$
for large arguments],
\begin{equation}
\bar{v}=\sqrt{\frac{8kT}{\pi m}},
\label{idealgas}
\end{equation} 
which is the mean speed of a molecule of an ordinary nonrelativistic gas (see e.g. Ref. \cite{lan80}). Thus, as $T$ approaches the absolute zero,
the massive photons disappear from the cavity [cf. Eq. (\ref{number2})], and in doing so they come to rest. The plot in Fig. \ref{pressaobessel} illustrates these facts.
%%%%%%%%%%%%%%%FIGURE%%%%%%%%%%%%%%%%%%%%%%%
%%%%%%%%%%%%%%%%%%%%%%%%%%%%%%%%%%%%%%%%%%%%%%%%%%%

\section{Radiance}

As $N$ in Eq. (\ref{number}), the internal energy 
$U$ can also be obtained by using Eq. (\ref{distribution}),
\begin{eqnarray}
&&\frac{U}{V}=\int \frac{d^3 p}{h^3}\epsilon f({\bf p})=
\frac{8\pi}{h^3}\int_{0}^{\infty}\- dp\ \frac{p^2\epsilon}{e^{\beta\epsilon}-1}.
\label{uenergy}
\end{eqnarray}
In order to make contact with the corresponding Planckian formula,
it is convenient to consider $\epsilon=\hbar\omega$ and to rewrite
Eq. (\ref{uenergy}), noticing  Eq. (\ref{energy}), resulting in
\begin{equation}
\frac{U}{V}=\int_{mc^2/\hbar}^{\infty}
d\omega\ u(\omega,T,m),
\label{massivephoton}
\end{equation}
where
\begin{equation}
u(\omega,T,m):=\frac{\hbar}{\pi^2 c^3}
\frac{\omega^3}{e^{\beta \hbar\omega}-1}
\sqrt{1-(mc^2/\hbar\omega)^2}.
\label{planck}
\end{equation}
By setting $m=0$ in Eqs. (\ref{massivephoton}) and (\ref{planck})
the  Planckian result emerges as expected. [It should be remarked that
an expression for $U$ as that for $N$ in Eq. (\ref{number2}) is also available but it will not be needed here.]

The energy density $U/V$ as given by Eqs. (\ref{massivephoton}) and (\ref{planck}) is in agreement with Ref. \cite{tor85}. According to 
Ref. \cite{tor85} the radiance ${\cal R}$ (which is defined as the amount of 
energy radiated per second per unit of area of an opening on the wall
of the cavity) is related to the energy density as,
\begin{equation}
{\cal R}=\frac{c}{4}\frac{U}{V}.
\label{toresult}
\end{equation} 
In the following it will be shown that Eq. ({\ref{toresult}})
is incorrect if the photon has a nonvanishing mass.

Consider a small opening on the wall of the cavity. The power
radiated  per unity of area can be obtained by calculating  the flux of $(d\Omega/4\pi)u{\bf v}$ through the surface of the opening 
(see e.g. Ref. \cite{car05}) taking into
account all possible values of $\epsilon=\hbar\omega$ and
all possible directions of flight starting from inside the cavity: 
\begin{equation}
{\cal R}=\int_{mc^2/\hbar}^{\infty}
d\omega\ (uv/4),
\label{radiance}
\end{equation}
with $v$ and $u$  given in Eqs. (\ref{speed}) and (\ref{planck}). It should be noticed that $uv/4$
in Eq. (\ref{radiance}) came about by solving
$$\int\frac{d\Omega}{4\pi} uv \cos\theta,$$
where the integration considers only one hemisphere  and $\theta$ is the angle between the straight line normal to the opening and the straight line associated with
photon's flight out of the cavity. Clearly, Eq. (\ref{toresult})
follows from Eq. (\ref{radiance}) only for massless photons
[cf. Eqs. (\ref{speed}) and (\ref{massivephoton})].

In a manner similar to that which led to Eq. (\ref{geometricseries}), the use of the
geometric series allows one to evaluate the integration in
Eq. (\ref{radiance}),
\begin{equation}
{\cal R}=\frac{3}{2(\pi c)^2 \hbar^3}
(k T)^4\left[Li_{4}(e^{-x})
+ x\ Li_{3}(e^{-x})+ \frac{x^2}{3}Li_{2}(e^{-x})
\right].
\label{r}
\end{equation}
It is instructive to consider the behavior of Eq. (\ref{r})
for small mass [more precisely when $x\ll 1$, cf. Eq. (\ref{number2})], namely
\begin{eqnarray}
&&
{\cal R}=\frac{\pi^2}{60\hbar^3 c^2}(k T)^4
\left[1-\frac{5}{2\pi^2}\left(\frac{mc^2}{kT}\right)^2\right],
\hspace{1.5cm}
\frac{mc^2}{kT}\ll 1,
\label{corrections}
\end{eqnarray}
where one recognizes the first term in the expression for 
${\cal R}$ in Eq. (\ref{corrections}) as the Stefan-Boltzmann law of blackbody radiation. Now, noticing Eq. (\ref{poly}), for a nonvanishing mass and at very low temperatures Eq. (\ref{r}) yields
\begin{eqnarray}
&&
{\cal R}=\frac{1}{2\hbar^3}\left(\frac{mc}{\pi}\right)^{2}(k T)^2\ e^{-mc^2/kT},
\hspace{1.5cm}
\frac{mc^2}{kT}\gg 1,
\label{corrections2}
\end{eqnarray}
showing that ${\cal R}$ fades away much more quickly
than its massless counterpart in the Stefan-Boltzmann law
as the absolute zero of temperature is approached.

\section{Final remarks}
Before closing, it should be pointed out that the second
term in the expression for ${\cal R}$ in Eq. (\ref{corrections})
is twice that in Ref. \cite{tor85} and therefore the amendments proposed
above do not improve much the possibility of measuring the mass of the photon  by considering the deviation from the Stefan-Boltzmann law in Eq. (\ref{corrections}) \cite{tor85}.
However, by progressively lowering the temperature, at some point the
radiance ${\cal R}$ for massive photons in Eq.  (\ref{r}) will differ radically
from the Stefan-Boltzmann law
[cf. Eq. (\ref{corrections2})]. 
A word of caution is due regarding this remark though. In the context of 
quantum field theory at finite temperature, 
one learns that the use of the distribution in Eq. (\ref{distribution}) to 
address radiation in a cavity of finite volume is not always a reliable approach.
Rougly speaking, if the smallest dimension $a$ of the cavity is such that $kTa/\hbar c$
is not big enough (ideally infinite), quantum vacuum cannot be ignored, and it will play a
role, resulting that Eq.  (\ref{r}) no longer can be trusted \cite{bro69}.

\section*{Acknowledgements}
\hspace{0.1cm}
This work was partially supported by the ``Coordena\c{c}\~{a}o de Aperfei\c{c}oamento de Pessoal de N\'{\i}vel Superior'' (CAPES) and the  ``Funda\c{c}\~{a}o de Amparo \`{a} Pesquisa do Estado de Minas Gerais'' (FAPEMIG).
%research agencies CNPq and FAPEMIG.
%

%\section*{References}

%\end{multicols}

\end{document}